\pdfoutput=1
\documentclass{jors}
\usepackage{graphicx}
\usepackage{circledsteps}
\pgfkeys{/csteps/inner color=white}
\pgfkeys{/csteps/fill color=black}
\usepackage{textcomp}
\usepackage[numbers]{natbib}
\usepackage{amsmath}
\usepackage[frozencache,cachedir=.]{minted}
\definecolor{mintedbg}{rgb}{0.95,0.95,0.95}
\usepackage[utf8]{inputenc}
\DeclareUnicodeCharacter{2082}{\textsubscript{2}}
\DeclareUnicodeCharacter{03C1}{$\rho$}

\definecolor{mygray}{gray}{0.6}

\usepackage{booktabs}
\newcommand*{\doi}[1]{DOI: \href{http://dx.doi.org/#1}{\textcolor{blue}{#1}}}
\newcommand{\wnbc}[1]{{\small\Circled{\textbf{\sffamily #1}}}}

\DeclareRobustCommand{\BibTeX}{{\normalfont B\kern-.05em{\scshape i\kern-.025em b}\kern-.08em \TeX}}
\begin{document}

{\bf Software paper for submission to the Journal of Open Research Software} \\

To complete this template, please replace the blue text with your own. The paper has three main sections: (1) Overview; (2) Availability; (3) Reuse potential. \\

Please submit the completed paper to: editor.jors@ubiquitypress.com

\rule{\textwidth}{1pt}

\section*{(1) Overview}

\vspace{0.5cm}

\section*{Title}
Yamdb: easily accessible thermophysical properties of liquid metals and molten salts

\section*{Paper Authors}

1. Weier, Tom\\
2. Nash, William\\
3. Personnettaz, Paolo\\
4. Weber, Norbert
\section*{Paper Author Roles and Affiliations}
1. Staff scientist, Institute of Fluid Dynamics, Helmholtz-Zentrum Dresden-Rossendorf, Germany\\
2. PostDoc, Institute of Fluid Dynamics, Helmholtz-Zentrum Dresden-Rossendorf, Germany\\
3. PostDoc, Centre National d’Etudes Spatiales, Paris, France;
Univ.~Grenoble Alpes, Univ.~Savoie Mont Blanc, CNRS, IRD,
Univ.~Gustave Eiffel, ISTerre, 38000 Grenoble, France; Institute of Fluid Dynamics, Helmholtz-Zentrum Dresden-Rossendorf, Germany\\
4. Staff scientist, Institute of Fluid Dynamics, Helmholtz-Zentrum Dresden-Rossendorf, Germany
\section*{Abstract}

Yamdb (Yet another materials data base) addresses the need to provide
thermophysical properties of liquid metals and molten salts in an
easily accessible manner. Mathematical relations describing material
properties - usually determined by experiment - are taken from the
literature. Equations and their coefficients are stored
separately. The former can be implemented in any programming language
(Python and Go in this case) and the latter are kept in YAML files
together with additional information (source, temperature range,
composition, accuracy if available, etc).
\section*{Keywords}

material properties; liquid metals; molten salts; YAML; Python; Go

\section*{Introduction}
Knowledge of materials properties is a necessary prerequisite for
a great number of scientific and engineering tasks, such as designing
apparatuses, estimating heat fluxes, ohmic losses and other quantities
and is indispensable for many numerical simulations, e.g., in
computational fluid dynamics. Correspondingly many material databases
exist, typically built to address the needs of a specific field. We
concentrate here on liquid metals and fused salts mainly for their
application in electrochemical energy storage, i.e., in liquid metal
and molten salt batteries \cite{KimBoysenNewhouseEtAl:2013,
  KelleyWeier:2018}.  Yamdb was written to offer rapid and easy access
to a number of material properties in the liquid state at standard
pressure. Typically, the substances useful in this field possess
melting temperatures significantly above room temperature.

The scarcity of data on the thermophysical properties of liquid
metals compared to, e.g., hydrocarbons is due to the experimental
difficulties in obtaining accurate measurements at high temperatures
\cite{IidaGuthrie:2015}. The same holds true for fused
salts. Nevertheless, there are several carefully edited compilations
of thermophysical properties of liquid metals
\cite{Ohse:1985,IidaGuthrie:1988,IidaGuthrie:2015} available in text.
Janz's collection \cite{Janz1967, Janz1988} of molten salt properties from the literature published until about 1988 is a seminal
work still held in high esteem
\cite{Serrano-LopezFraderaCuesta-Lopez2013} even if these data are no
longer considered Standard Reference Data (SRD) by the National
Institute of Standards and Technology (NIST) \cite{Janz1992}.

The challenges of building an up-to-date electronic molten salts
database containing critically evaluated data are vividly described by
Gaune-Escard and Fuller \cite{Gaune-EscardFuller2000,
  FullerGaune-Escard2003}. Assessing data quality and accuracy of
reporting is a task that should not be underestimated and that entails
considerable resource requirements \cite{Chiricoetal2013}. Even
defining a standardized universal data format for thermophysical
properties has proved to be a major undertaking \cite{SemerjianBurgess2022}.



Our aim with Yamdb is much more modest and focused on molten metals
and fused salts: we collect correlations, based on experimental
findings, describing temperature dependent thermophysical properties
from the literature and make them easily accessible from (interactive)
programming environments. Value judgments were not
attempted. Collecting clean data from the literature is thereby the
main work. Though it is seemingly trivial, errors easily creep in or
are already present in the original sources
\cite{PfeifKroenlein2016}. Using multiple sources for identical
properties wherever possible helps to discover errors. Since such
errors are unavoidable and often difficult to detect, correct numbers
(or coefficients) are of greater value and have considerably longer
lifetimes than the concrete implementation of equations. Both should
therefore be stored separately. To keep this asset, an easily
editable, flexible, well documented data format with a permissible
license and widespread adoption is needed. Longevity is vital as well
but difficult to assess.

We decided here for YAML \cite{Ben-Kikietal2021, Hinsen2012,
  Hinsen2018} (YAML Ain't Markup Language\texttrademark{}) because it fulfills most of the aforementioned criteria
and is a core component of a sufficient number of popular software
packages which gives hope for sustained support. In our opinion, YAML strikes
a good balance between the complexity of the database format and flexibility.

\begin{table}
  {\small
  \begin{tabular}{lllll}
    \toprule
    Name    & Programming & Ref.                  & License \\
    \midrule
    Cantera & C++/Fortran/Python & \cite{Goodwinetal2022} & BSD \\
    DDB & &\cite{Onkenetal1989} & commercial & \\
    MSTDB-TP & C++ & \cite{Terminietal2023} & copyrighted/license-free \\
    NIST alloy data & web-based & \cite{Wilthan2019} & public domain\\
    NIST Molten  & & & \\
    \hspace*{0.7em}Salts Database & &\cite{ChaseSauerwein1993} & commercial \\ 
    PYroMat & Python   & \cite{Martinetal2022} & GNU GPL v3.0 \\
    pyThermoML & Python & \cite{Gueltigetal2022} & BSD \\
    REFPROP & Fortran & \cite{Lemmonetal2018} & commercial\\
    Thermo  & Python & \cite{BellContributors2016} & MIT \\
    ThermoData Engine & C++ & \cite{Frenkeletal2005} & commercial \\
    ThermoFun & C++/Python & \cite{Mironetal2023} &  LGPL\\
    ThermoML  & XML/JSON & \cite{Riccardietal2022} & open \\ 
    TPDS-web & web-based &\cite{AIST2006, Yasudaetal2020} & copyrighted/free of charge\textsc{}\\
    \bottomrule
  \end{tabular}
  }
  \caption{\label{tab:OtherCodes}Selected software and databases for thermodynamic properties.}
\end{table}

As mentioned above, over the years a whole series of databases and
tools for using thermophysical data have been developed. For a small
subset see Tab.~\ref{tab:OtherCodes}. The table collects software and
databases from different fields. Driven by demand from the chemical
industry, several well established and comprehensive databases with a
focus on hydrocarbons exist, e.g., the Dortmund Data Bank (DDB)
\cite{RareyGmehling2009}, the ThermoData Engine (TDE)
\cite{Frenkeletal2005} and REFPROP \cite{Huberetal2022} from
NIST. They often contain sophisticated modeling capabilities for
thermochemical and transport properties especially for
mixtures. Cantera \cite{Goodwinetal2022}, PYroMat
\cite{Martinetal2022}, Thermo \cite{BellContributors2016}, and ThermoFun
\cite{Mironetal2023} are examples of open source software belonging
to this category.
The NIST Molten Salts Database \cite{ChaseSauerwein1993} is a DOS
based program that offers easy access to the collection of correlation
equations for the density, dynamic viscosity, electric conductivity,
and surface tension of molten salts collected by Janz and co-workers
\cite{Janz1988}. Recently, NIST made the underlying coefficients and
equations electronically available \cite{Janz1992}. The Molten Salt
Thermal Properties Database (MSTDB) \cite{Terminietal2023} with a
thermochemical (MSTDB-TC) and a thermophysical branch (MSTDB-TP) is
being actively developed and focuses on the properties of molten salts
suitable for the use in molten salt reactors. ThermoML is primarily
a data description in Extensible Markup Language (XML) standardized by
the International Union of Pure and Applied Chemistry (IUPAC)
\cite{Frenkeletal2006, Frenkeletal2011}. NIST provides an archive of material property
values stored in their ThermoML Archive \cite{Riccardietal2022} with
the possibility to access the data in JavaScript Object Notation
(JSON) as well. It includes several sets of measured molten salt
property values. PyThermoML \cite{Gueltigetal2022} wraps ThermoML data
sets into Python data structures but does not yet - at the time of writing -
implement the specification in its entirety. Material
properties of molten metals and alloys are available as sets of
experimental data from NIST \cite{Wilthan2019} and in a somewhat
smaller quantity from the Network Database System for Thermophysical
Property Data (TPDS-web) developed by the Japanese National Institute
of Advanced Industrial Science and Technology (AIST) \cite{AIST2006,
  Yasudaetal2020}. This database also contains data on molten salts.


So far Yamdb has been used in a number of publications
\cite{Personnettazetal2022,Leeetal2023,Duczeketal2023a} on liquid
metal and molten salt batteries as well as on interfacial
instabilities \cite{Weieretal2020,Noreetal2021} in liquid metal
systems with a miscibility gap.

\section*{Implementation and architecture}

\begin{figure}
  \includegraphics[width=0.637\textwidth]{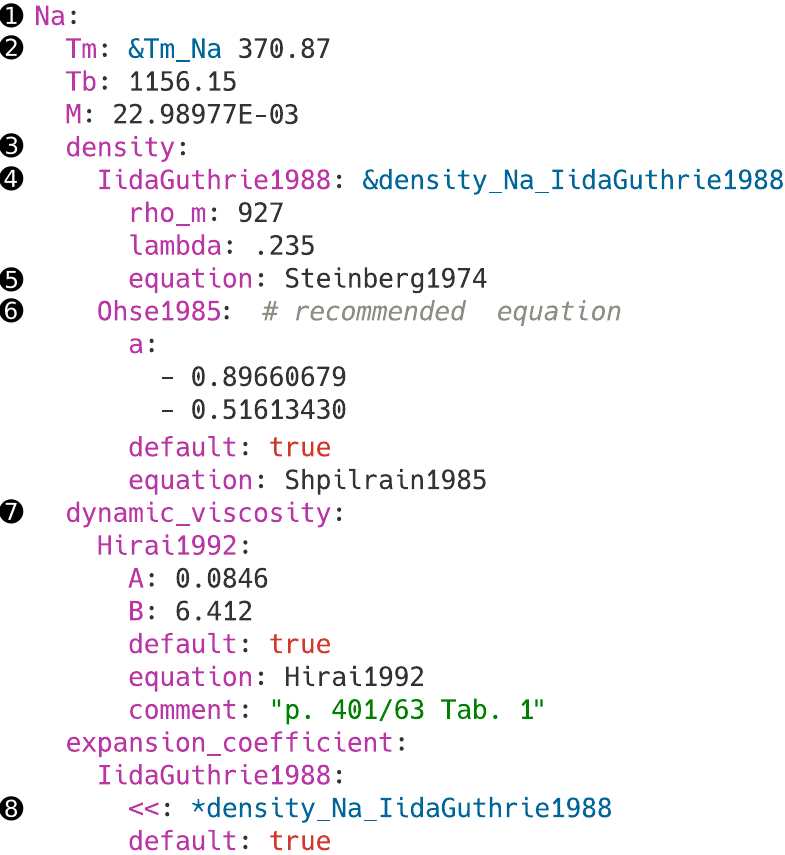}
  \caption{\label{fig:metal_mod_yaml}Modified snipped from the \texttt{metals.yml} database file showing sodium properties. Please see the text for further explanations.}
\end{figure}

As mentioned above, the coefficients for the equations expressing the various properties depending on temperature (and concentration) are
stored in YAML files. These files contain a nested structure of block
mappings \cite{Ben-Kikietal2021} so that the symbol of a substance
(element or mixture, `\texttt{Na}' in the example in
Fig.~\ref{fig:metal_mod_yaml}) constitutes the key of the root node
\wnbc{1}. This root node has a number of
mandatory child nodes: melting temperature \texttt{Tm}
\wnbc{2}, boiling temperature \texttt{Tb}, and molar
mass \texttt{M}. The temperatures are given in K, the molar mass in kg/mol.
Thermophysical properties (\texttt{density} \wnbc{3},
\texttt{dynamic\_viscosity} \wnbc{7}, and
\texttt{expansion\_coefficient}) are sibling nodes of the aforementioned ones
and might or might not be present depending on data availability. The
property node with the key \texttt{density} \wnbc{3}
has two child nodes \texttt{IidaGuthrie1988} \wnbc{4}
and \texttt{Ohse1985} \wnbc{6} the keys of which are
named according to the references for the original coefficient data:
\cite{IidaGuthrie:1988, Ohse:1985}.
The child node with the key \texttt{equation}
\wnbc{5} designates the equation used for the
density calculations, in this case
\begin{equation}
  \label{eq:Steinberg1974}
  \rho = \rho_\text{m} - \lambda (T - T_\text{m})
\end{equation}
derived from \cite{Steinberg:1974} and
\cite{IidaGuthrie:1988}\footnote{Note that the sign of $\lambda$ is
  inverted with respect to the original definition.}. $\rho$ is the
density in kg/m\textsuperscript{3} at the temperature $T$ in K, while
$\rho_\text{m}$ is the density at the melting point $T_\text{m}$ and
$\lambda$ the temperature dependence of the liquid density in kg/(m\textsuperscript{3} K).

If the density equation to be used is not explicitly specified
however, it would be calculated according to the data from \texttt{Ohse1985}
and the equation \texttt{Shpilrain1985} because the \texttt{Ohse1985} block
\wnbc{6} contains the \texttt{default} key with the
value \texttt{true}. A convenient feature of YAML are the
`anchor'(\&)-`alias'(*) pairs that allow direct access to the
coefficients of the density equation according to \texttt{IidaGuthrie1988}
\wnbc{4} by the corresponding equation for the
expansion coefficient \wnbc{8} without
duplicated input. The \texttt{default: true} key-value pair is added to the  
received coefficients.

The presence of several keys in the YAML databases are required: \texttt{Tm},
\texttt{Tb}, \texttt{M} for each substance, the necessary parameters for the
respective equations, one and only one \texttt{default} per property,
\texttt{equation} per equation record, and \texttt{reference} if the equation record's
key cannot be mapped to an entry in the references database. If
available, ranges of temperature (\texttt{Tmin}, \texttt{Tmax}) or fraction in mol\%
(\texttt{xmin}, \texttt{xmax}) for which the correlations are valid are included as
well as information about the accuracy (\texttt{uncertainty}) and a \texttt{comment}
for remarks and explanations. Further keys can be added as needed.
If necessary, additional methods can be implemented that make use of
the content referenced by these keys.

As mentioned above, substances are addressed by their chemical
symbols. Ordering of these symbols in compounds is alphabetical. In
the case of salts, we follow the convention used by Janz
\cite{Janz1988} and start the formula with the cationic species,
e.g., `NaCl' and not `ClNa' as often found in the newer literature,
e.g., \cite{Haynesetal2016}. Components of mixtures are divided by a
dash and ordered alphabetically by compound. The popular mixture of
lithium chloride and potassium chloride would therefore be found under
`KCl-LiCl'.

Since YAML is a human-friendly text format, the database files can
be conveniently edited in a number of editors and integrated
development environments (IDEs) with YAML support, e.g., Emacs with
yaml-imenu and yafolding, Geany, PyCharm, and Vim.

Python was chosen for the primary implementation because of its well
established role as today's lingua franca of scientific programming
and the truly excellent REPL (Read-Eval-Print Loop) IPython. IPython
enables interactive, exploratory work \cite{Perezetal2011} and thus
easy and rapid access to the available property methods of the
dynamical generated substance objects of Yamdb. 

There are of course valid reasons to prefer other languages,
e.g., escaping frequent deprecation and breaking changes (even in the
core language and libraries) \cite{Hinsen2019,Veytsman2021} or the
desire to use less electricity when performing computations
\cite{Zwart2020}.

We chose the Go programming language \cite{DonovanKernighan2015} as a
case in point for an alternative implementation of a subset of the
Yamdb features in the standalone command-line program Goma (Go
materials database). Go has a focus on simplicity and long term
maintenance and promises compatibility \cite{Coxetal2022} at the
source level for the lifetime of the Go 1 specification. In addition,
Go offers, among other things, easy portability by supporting a
variety of operating systems and processor architectures via native as
well as cross compiling. However, the discussion below will mainly
focus on the Python implementation Yamdb.

To minimize maintenance effort and increase the longevity of software,
it is crucial to reduce dependencies as much as possible and to select
components with proven track records \cite{Hinsen2019}. We have tried to
heed this advice and aimed for minimal dependencies besides the
essential one on a YAML parser. Other than that, only NumPy and an
(optional) \BibTeX{} parser are required third party libraries. The
remaining external modules are part of the Python Standard
library. There is a long-lasting discussion in the Python community
\cite{Cannonetal2016} on the right place for a YAML parser with voices
\cite{Zaczynski2022} calling it ``Python's missing battery''. Time
will tell, whether YAML will be found worthy of inclusion one day. NumPy
is used in the equations for the simple reason that it allows
evaluation of thermophysical properties for a vector of
temperature values with one function call. NumPy
\cite{vanderWaltetal2011,Harrisetal2020} is a fundamental package for
scientific computing in Python and therefore not likely to disappear
anytime soon. In case of emergency (the release of NumPy 2.0 is nigh),
all calls to NumPy should be replaceable by those to the Python Math
Library at the expense of loosing the array computing capabilities. To
retain the focus and keep the library a manageable size, other
functionality like plotting via Matplotlib is deliberately excluded.

\begin{figure}
  \centering
  \includegraphics[width=0.7\textwidth]{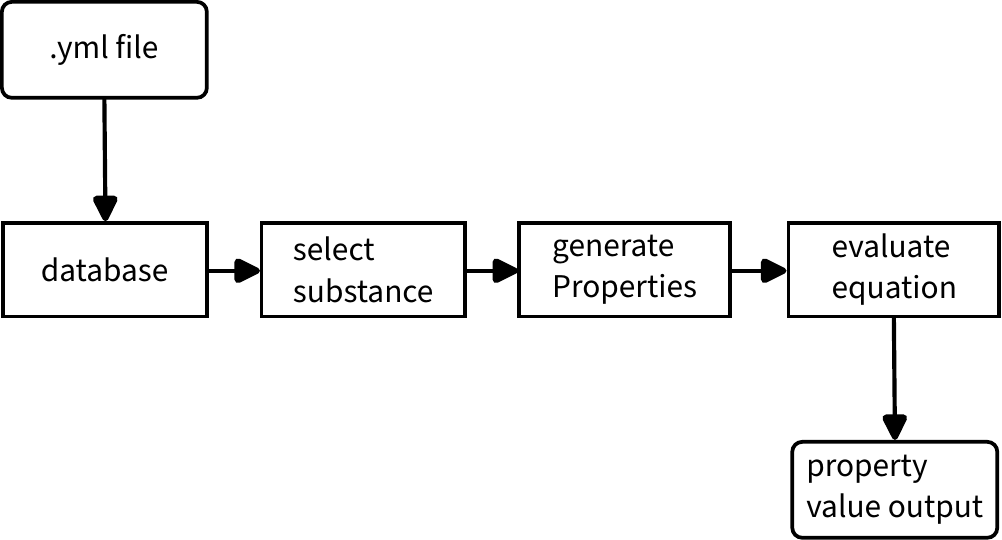}
  \caption{\label{fig:architecture}Overall program flow when using
    Yamdb. Please see the text for further explanations.}
\end{figure}

The overall program flow is shown in Fig.~\ref{fig:architecture}
and followed in the listing below

\begin{minted}[linenos,bgcolor=mintedbg,numberblanklines=false]{python}
import os
from importlib import resources as ir
from yamdb import yamdb as ydb

metals_yml = os.path.join(ir.files('yamdb'), 'data', 'metals.yml')
sdb = ydb.SubstanceDB(metals_yml)
Na_dict = sdb.get_substance('Na') 
Na = ydb.Properties(Na_dict)
T = Na.Tm
rho = Na.density(T)

print('Na density at %f K is %f kg/m³' % (T, rho))
\end{minted}

After loading Yamdb first, a database object has to be generated from
a supplied YAML file (line 5). The main component of this object is a
dictionary (\texttt{sdb.materials}; internal use only, not present in
the listing) of dictionaries corresponding to the structure of the
loaded YAML file. The next step on line 6 is the selection of a
sub-dictionary \texttt{Na\_dict} whose key `Na' in
\texttt{sdb.materials} corresponds to the symbol of the substance to
be investigated. In this case it is the element sodium. From this
sub-dictionary, the property object \texttt{Na} is dynamically built
by a call to \texttt{yamdb.Properties}. This is done by assigning
values from the sub-dictionary to the three required constant
properties melting and boiling point and molar mass (\texttt{Tm},
\texttt{Tb}, \texttt{M}). Furthermore, for the properties that are
implemented in the sub-modules of \texttt{yamdb.properties} and that
are present in the sub-dictionary \texttt{Na\_dict}, methods are
dynamically generated. They select a default equation per property and
allow the user to address the non-default equations by the keyword-argument
\texttt{source}. If one would like to know the density of sodium
according to \texttt{IidaGuthrie1988} instead of the default
\texttt{Ohse1985} (see Fig.~\ref{fig:metal_mod_yaml}), one would call
\mintinline{python}{rho = Na.density(T, source='IidaGuthrie1988')}
instead of the simpler line 9. The property object \texttt{Na} now
allows one to conveniently access all available property methods by
tab-completion at the command line of IPython and recent Python
interpreters with access to the readline library. A number of
additional methods exist that allow one to extract different kinds of
information from the properties object such as listing all available
sources for one property (\texttt{get\_source\_list}) or getting the
reference key for a certain source (\texttt{get\_reference}).

A more convenient variant of the procedure described above is offered
by the function \texttt{get\_from\_metals} in line 2 of the listing
below.

\begin{minted}[linenos,bgcolor=mintedbg,numberblanklines=false]{python}
from yamdb import yamdb as ydb

Na = ydb.get_from_metals('Na')

print('Na density at %f K is %f kg/m³' % (Na.Tm, Na.density(Na.Tm)))
\end{minted}

It directly accesses the \texttt{metals.yml} database included in the
Yamdb distribution and directly returns the corresponding Property
object. The first call to \texttt{get\_from\_metals} parses the YAML
file, and all subsequent calls for other substances use a cached version
of the corresponding dictionary.

Mixtures are handled slightly differently because they are not only
defined by their components but also by their
mole-fractions. Therefore objects of the \texttt{Mixture\-Properties} class

\begin{minted}[linenos,bgcolor=mintedbg,numberblanklines=false,mathescape=true]{python}
from yamdb import yamdb as ydb

CaCl2_NaCl = ydb.get_from_salts('CaCl2-NaCl')

T = 1020
st = CaCl2_NaCl.composition['20-80'].surface_tension(T)

print('CaCl₂-NaCl 20-80 surface tension at %d K is %4.3f N/m' %
 (T, st))

rho = CaCl2_NaCl.composition['range'].density(1090, 15)
# ρ = 1641 kg/m³

rho = CaCl2_NaCl.composition['15-85'].density(1090)
# ρ = 1635 kg/m³
\end{minted}

are (transparently) generated by the function
\texttt{get\_from\_salts} if a hyphen is detected in the substance
name. Properties are now methods of substance and composition as shown
in line 4 of the listing above. Compositions in the dictionary keys
are quoted in mol\%, the first value (20) is for the first component
(CaCl\textsubscript{2}), the second (80) for the second (NaCl) and so
forth. Composition keys are kept as short as possible, trailing zeroes
are removed and values are rounded to one digit after the decimal
point. If correlation equations for a continuous range of compositions
are available, the composition dictionary key is `\texttt{range}' as
in line 7. The property function (density in this case) requires then
the fraction in mol\% as a second argument. As can be seen from the
comment lines 8 and 10, density values for 1090\,K and
CaCl\textsubscript{2}-NaCl 15-85 from the two different correlation
functions are quite close.

It is essential to check the original sources before using material
properties derived from Yamdb in calculations, and publications should
cite the original sources. To facilitate these tasks, Yamdb contains a
module \texttt{yamref} and a \BibTeX{} database with all references
\texttt{references.bib}. \BibTeX{} was chosen because it can be
considered a standard format in scientific publishing, interacts
readily with \LaTeX{} and has a long history without breaking changes. A
number of \BibTeX{} parsers are available as third party Python
packages. Yamdb currently uses BibtexParser \cite{Weiss2023} that
depends on PyParsing \cite{McGuire2023}. At the time of writing,
BibtexParser undergoes a major version upgrade that entails breaking
changes to the application programming interface (API). In future,
Yamref might switch to an alternative solution for parsing \BibTeX{}
files but without changes to the user interface. For now, a simple
backup solution is provided in form of a YAML file mirroring the
content of the \BibTeX{} file. Access to the YAML file with references
is provided by functions implemented in Yamdb's yamdb module. Identity
of the information content of both versions (\BibTeX{} and YAML) is
covered by an integration test, see section `Quality control'.

A simple case of looking up the reference for a density value is
demonstrated in the listing below:

\begin{minted}[linenos,bgcolor=mintedbg,numberblanklines=false,mathescape=true]{python}
from yamdb import yamdb as ydb
from yamdb import yamref as yrf

Ga = ydb.get_from_metals('Ga')

rho = Ga.density(273.15 + 50)

yrf.get_from_references(Ga.get_default_source('density'))
# 'Iida, Takamichi and Guthrie, Roderick I. L (1988)
# The Physical Properties of Liquid Metals.
# Clarendon Press. Oxford.'

# alternatively use (references.yml):
ydb.get_from_references((Ga.get_default_source('density'))
# 'Iida, T., Guthrie, R.I.L., 1988. The physical properties
# of liquid metals. Clarendon Press, Oxford.'
\end{minted}

\texttt{get\_default\_source('density')} returns the source key of the
default source (\wnbc{6} in Fig.~\ref{fig:metal_mod_yaml}). It is
\texttt{IidaGuthrie1988} for Ga that gets expanded to the full
citation by the call to \texttt{yamref.get\_from\_references}. If
different sources from the same reference are present, the sources'
keys, e.g., \texttt{Ohse1985Rec} and \texttt{Ohse1985Exp}, are derived
from but do not directly correspond to the reference. In such a case,
an intermediate call to \texttt{yamdb.get\_reference} is necessary.

In contrast to the module Yamdb, Goma is a single program whose output
is controlled by command-line switches. The source code is divided
into a library of material property functions  and
the main program. It basically evaluates the arguments passed to Goma
and directs them to the proper functions for property evaluation in a
\texttt{switch} statement. Goma depends on the Go Standard library and
go-yaml v.2  \cite{Niemeyer2020}. Thanks to the versatile dependency 
management tools present in the Go-ecosystem, Goma includes the go-yaml
v.2 source via

\begin{minted}[bgcolor=mintedbg,numberblanklines=false]{bash}
> go mod vendor
\end{minted}

Goma calling conventions are described by a help screen (\texttt{goma
  -h} or \texttt{goma -help}, command line flags follow Plan9
standards \cite{ButcherFarina2016}) and a man page. To get the density
of Ca at 1130\,K enter

\begin{minted}[bgcolor=mintedbg,numberblanklines=false]{bash}
> goma -S Ca -T 1130 -density
# property  Temp source            value  unit
#           K                             
# --------- ---  --------          -----  ----
# density   1130 IidaGuthrie2015   1361.7 kg/m³
# density   1130 Shpilrainetal1980 1351.2 kg/m³
# --------- ---  --------          -----  ----
\end{minted}

at the command line. The available information is returned as a
table. Resolving references is not done internally, only the citation
keys are provided in the column `source'. Should the user have
access to \LaTeX{}  and pandoc \cite{MacFarlane2023}, a shell script
\texttt{gen\_report.bash} supplied with Goma can be used to generate
\begin{figure}
  \centering
  \includegraphics[width=0.86\textwidth]{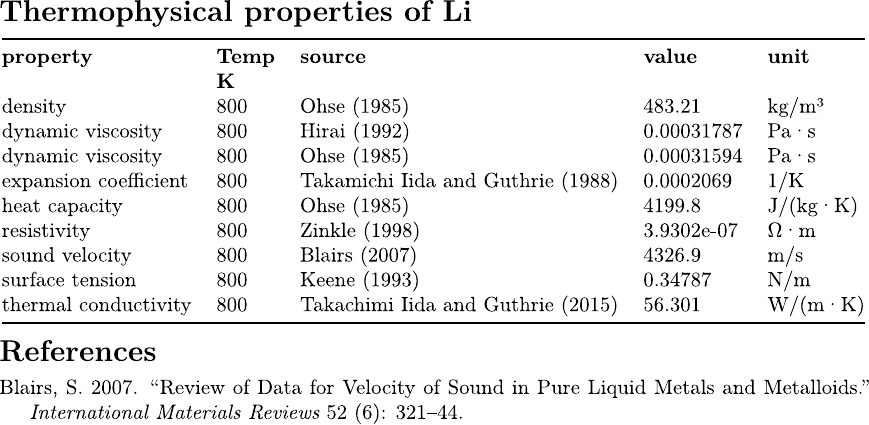}
  \caption{\label{fig:report}Script-generated report on the
    thermophysical properties of Li at 800\,K using Goma
    (shortened). Formatting and bibliography require pandoc and
    \LaTeX.}
\end{figure}
an overview of all thermophysical properties available for a
substance at a given temperature. A (shortened) example of the
script's output for Li at 800\,K is shown in Fig.~\ref{fig:report}.

\section*{Quality control}


\begin{figure}
  \includegraphics[width=0.8\textwidth]{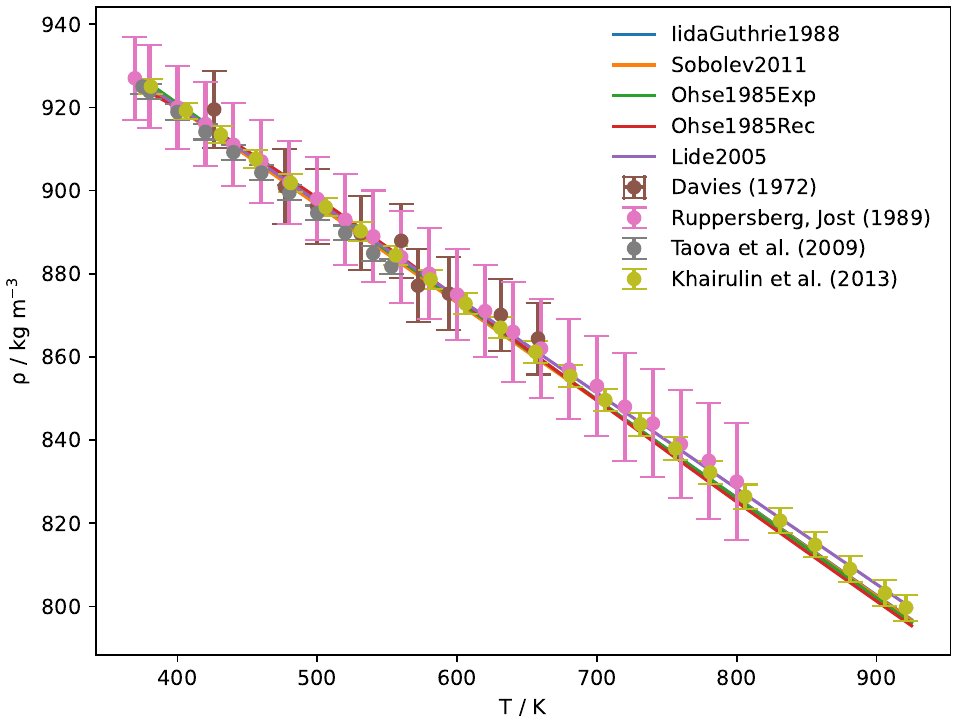}
  \caption{\label{fig:sodium_density}Sodium density according to
    correlation equations from Yamdb (solid lines) compared to
    experimental data provided via the NIST Alloy data web application
    \cite{Wilthan2019, PfeifKroenlein2016, Wilthanetal2017}. The original
    data are from Davies (1972) \cite{Davies1972}, Khairulin et
    al.~(2013) \cite{Khairulinetal2013}, Ruppersberg and Jost (1989)
    \cite{RuppersbergJost1989}, and Tavo et al.~(2009) \cite{Taovaetal2009}}.
\end{figure}

Yamdb uses pytest for quality control. The single property equations
are fully covered by unit tests. They check the correct implementation
of the equations by calculating a property value for a predetermined
temperature with a defined set of coefficients.

Completeness of the references and instantiation of objects are
examined by integration tests. This includes ensuring that all
citation keys of references.bib are present in references.yml.

Plausibility of the temperature dependency of the properties is tested
by comparing (if available) different equations for the same property
in one plot. Fig.~\ref{fig:sodium_density} shows an example for the
density values of sodium in the temperature range between melting and
boiling point. As can be seen, all correlation equations available
from Yamdb collapse fairly well and  are in the range of the error margins
of the most recent measurements by Khairulin et
al.~\cite{Khairulinetal2013}. Measured values included in
Fig.~\ref{fig:sodium_density} were taken from the NIST alloy data
\cite{Wilthan2019}. The \texttt{metals.yml} database can be checked by
a script \texttt{check\_metals.py} delivered with Yamdb that produces a
large number of plots comparing the temperature dependent
thermophysical properties for each metal between different sources.

An example for molten salt mixture properties is given in 
\begin{figure}
  \includegraphics[width=0.8\textwidth]{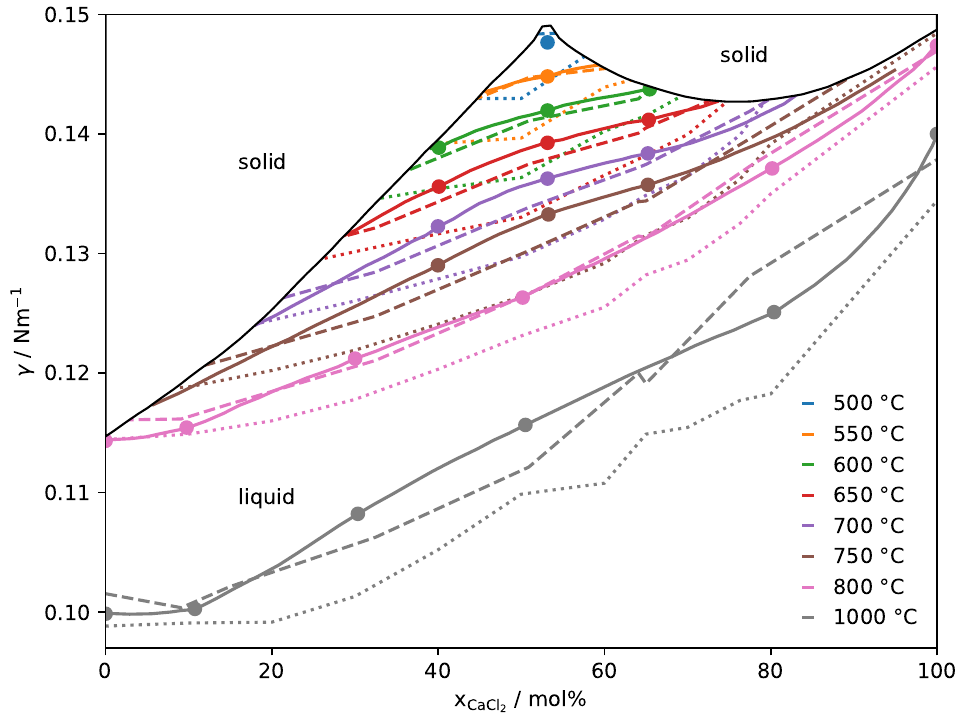}
  \caption{\label{fig:CaCl2-NaCl-surface_tension}Surface tension of
    CaCl\textsubscript{2}-NaCl mixtures versus composition for
    different temperatures adapted from
    \cite{AddisonColdrey1960}. Correlation expressions (dotted and
    dashed lines) from Yamdb compared to measured data by Addison,
    Coldrey \cite{AddisonColdrey1960} (dots and solid lines). Dotted
    lines result from evaluating the equations by \cite{Janzetal1975}
    and dashed lines from those of \cite{Janz1988}. As expected, the
    latter are identical to the values by \cite{Janz1992}.}
\end{figure}
Fig.~\ref{fig:CaCl2-NaCl-surface_tension} adapted from
\cite{AddisonColdrey1960}. It shows the surface tension of liquid
CaCl\textsubscript{2}-NaCl mixtures for different temperatures over
the whole range of compositions in the liquid state.  In the white
areas designated `solid', the mixture is not a liquid.  Measured
values provided by Addison and Coldrey \cite{AddisonColdrey1960} (dots
and solid lines) are compared to equations reported by Janz et
al.~\cite{Janzetal1975} and Janz \cite{Janz1988}. Temperature is
encoded by color. It is apparent that the earlier reported equations
by Janz et al.~\cite{Janzetal1975} (dotted lines) miss the measured
values by a difference in surface tension $\gamma$ corresponding to a
temperature difference of about 50\,K. The newer equations from Janz
\cite{Janz1988} (dashed lines) fit much better, especially for the
lower and higher temperatures. The values computed with the
coefficients from the database by Janz \cite{Janz1992} are identical
to the ones from Janz \cite{Janz1988} (dashed lines).

It should be emphasized that measurements of the thermophysical
properties of reactive liquid media at elevated temperatures are far
from trivial and very demanding \cite{MillsLee2001}. Certain
differences in the reported and calculated values are therefore
expected and simply reflect the experimental challenges.

The YAML databases are checked for correct syntax by using yamllint.

For the time being, Goma's results for predefined sets of temperature
values are simply compared to Yamdb's output.

\section*{(2) Availability}
\vspace{0.5cm}
\section*{Operating system}
\textbf{Yamdb}

An operating system capable of running a supported version of Python
is required. Yamdb is tested and developed on ArchLinux (rolling
release, 2023-11-03), and has been shown to run on OpenBSD 7.3 and 7.4
(requires installation of python3.11 from the ports tree), Ubuntu
22.04 with python3.11,
postmarketOS/Alpine Linux/SXMO, Armbian 23.5, Android 11 via termux,
as well as Microsoft Windows 10 with python3.12.

\textbf{Goma}

A Go compiler for the operating system and processor architecture is a
necessary precondition for compiling Go programs on a system.
Alternatively, cross-compilation can provide binaries for non-native
architectures and operating systems. Goma is tested and developed on
ArchLinux (rolling release, 2023-11-03), and has been shown to run on
OpenBSD 7.3 and 7.4,  Ubuntu
22.04 with python3.11, postmarketOS/Alpine Linux/SXMO, Armbian 23.5,
Android 11 via termux, as well as Microsoft Windows 10.

\section*{Programming language}

\textbf{Yamdb}

Python 3.11.3

\textbf{Goma}

Go 1.20.6

\section*{Additional system requirements}


No specific requirements.

\section*{Dependencies}


\textbf{Yamdb}

\begin{itemize}
\item BibtexParser version 1.4.0, \url{https://github.com/sciunto-org/python-bibtexparser}, BSD license
  \item NumPy version 1.25.1, \url{https://numpy.org/}, NumPy license
\item PyYAML version 6, \url{https://github.com/yaml/pyyaml}, MIT license
\end{itemize}

\textbf{Goma}

go-yaml version 2.4.0, \url{https://github.com/go-yaml/yaml/tree/v2}, Apache License Version 2.0

\section*{List of contributors}

Tanja Klopper (Helmholtz-Zentrum Dresden Rossendorf) - Development and testing\\
William Nash (Helmholtz-Zentrum Dresden Rossendorf) - Microsoft Windows port, testing, and language editing\\
Hirav Patel (Helmholtz-Zentrum Dresden Rossendorf) - Development and testing\\
Paolo Personnettaz (Helmholtz-Zentrum Dresden Rossendorf) - Development and testing\\
Norbert Weber (Helmholtz-Zentrum Dresden Rossendorf) - Material property collection and evaluation\\
Tom Weier (Helmholtz-Zentrum Dresden Rossendorf) - Design, development and testing, data collection\\

\section*{Software location:}



{\bf Archive Yamdb}

\begin{description}[noitemsep,topsep=0pt]
	\item[Name:] RODARE
	\item[Persistent identifier:] DOI: 10.14278/rodare.2549
	\item[Licence:] MIT
	\item[Publisher:]  Helmholtz-Zentrum Dresden-Rossendorf
	\item[Version published:] 0.3.0
	\item[Date published:] 05/11/23
\end{description}

{\bf Archive Goma}

\begin{description}[noitemsep,topsep=0pt]
	\item[Name:] RODARE
	\item[Persistent identifier:] DOI: 10.14278/rodare.2547
	\item[Licence:] MIT
	\item[Publisher:]  Helmholtz-Zentrum Dresden-Rossendorf
	\item[Version published:] 0.1.0
	\item[Date published:] 05/11/23
\end{description}



{\bf Code repository Yamdb}

\begin{description}[noitemsep,topsep=0pt]
	\item[Name:] Codebase.Helmholtz
	\item[Persistent identifier:] https://codebase.helmholtz.cloud/prosa/yamdb
	\item[Licence:] MIT
	\item[Date published:] 07/11/23
\end{description}

{\bf Code repository Goma}

\begin{description}[noitemsep,topsep=0pt]
	\item[Name:] Codebase.Helmholtz
	\item[Persistent identifier:] https://codebase.helmholtz.cloud/prosa/goma
	\item[Licence:] MIT
	\item[Date published:] 07/11/23
\end{description}



\section*{Language}

English.

\section*{(3) Reuse potential}

Yamdb may be used by anybody interested in the thermophysical
properties of liquid metals and molten salts. Such data are needed for
simulations in application areas like high temperature materials
processing, electrochemical engineering, and energy storage. Ease of
reuse and extensibility are central goals of Yamdb. It can be
incorporated as a library into simulation codes that feature a Python
interface. As demonstrated by Goma, the effort to transfer the
correlation equations into another programming language is relatively
low. Accessing the coefficients from the database files should be
trivial in most cases since several programming languages as Crystal
and Ruby ship YAML parsers as part of their standard libraries. For
most other languages well established third party parsers - often
based on the C library libyaml \cite{Project2021} - are
available. Direct access to temperature dependent thermophysical
properties could thereby be enabled for simulations in the language
best suited for the purpose.

Yamdb can be easily extended in a number of ways. In the simplest case
just by adding new correlation coefficients for existing substances
and already implemented equations. This requires only additions to the
YAML databases. New equations are easy to implement as well if
existing properties are to be extended. This requires only adding a
function with the standard interface

\begin{minted}[linenos, bgcolor=mintedbg,numberblanklines=false]{python}
def Shpilrain1985(Temp, coef=coef):
    r"""Return density according to Ohse (1985) p.453/454.
    ... (docstring shortened)
    """
    a = coef['a']
    rho = 0.0
    tau = Temp/1000.0
    for i, c in enumerate(a):
        rho += c * np.power(tau, i)
    rho *= 1000.0  # apparently needed to convert to kg/m³
    return rho
\end{minted}

to the package under \texttt{yamdb/properties} containing the
functions for the desired property, e.g., \texttt{density.py}. To make
copying of coefficients from the literature as fail-safe and easy as
possible, implementing the equation referred to in a paper is
preferred to complex coefficient conversions if nothing similar is
already available from the property modules.

If a new property is to be added, a corresponding module named after
the property (e.g., \texttt{dielectric\_constant.py} for the dielectric
constants) needs to be added to \texttt{yamdb/properties} and the
\texttt{module\_dict} in \texttt{yamdb/yamdb.py} has to be extended by
an entry for the new module (\texttt{'dielectric\_constant': None}).
In addition, the \texttt{class Properties} of \texttt{yamdb/yamdb.py}
requires a new private wrapper method for the extra property modelled
after:

\begin{minted}[linenos, bgcolor=mintedbg,numberblanklines=false]{python}
    def _density(self, Temp, *args, source=None):
        if source is None:
            source = self.get_default_source('density')
        return self._density_func_dict[source](Temp, *args)
\end{minted}

This can basically be done by replacing \texttt{density} with
\texttt{dielectric\_constant} in the listing above.

Adding new coefficients, functions, materials, and properties to Yamdb
is explained in detail in Yamdb's documentation.
 
Support for Yamdb and Goma is available by email to the authors of
this paper and by opening issues on Yamdb's or Goma's repository pages,
see paragraph `Software location/code repository'. Documentation and examples
are included in the software repositories. Contributing changes back
to the main repositories as pull requests or patches is encouraged
and highly welcome.

\section*{Acknowledgements}

Stimulating discussions with Michael Nimtz and Moritz Streb on the
needs for and the implementation of thermophysical materials databases
especially for liquid metals and molten salts were instrumental and
are highly appreciated. Paolo Personnettaz worked on the development
and testing of this project mainly during its stay at HZDR until
February 2022.

\section*{Funding statement}

This project has received funding from the European Union’s Horizon
2020 research and innovation programme under grant agreement No
963599.

\section*{Competing interests}

The authors declare that they have no competing interests.



\bibliographystyle{plainnat-doi}
\bibliography{references.bib}

\vspace{2cm}

\rule{\textwidth}{1pt}

{ \bf Copyright Notice} \\
Authors who publish with this journal agree to the following terms: \\

Authors retain copyright and grant the journal right of first publication with the work simultaneously licensed under a  \href{http://creativecommons.org/licenses/by/3.0/}{Creative Commons Attribution License} that allows others to share the work with an acknowledgement of the work's authorship and initial publication in this journal. \\

Authors are able to enter into separate, additional contractual arrangements for the non-exclusive distribution of the journal's published version of the work (e.g., post it to an institutional repository or publish it in a book), with an acknowledgement of its initial publication in this journal. \\

By submitting this paper you agree to the terms of this Copyright Notice, which will apply to this submission if and when it is published by this journal.

\end{document}